\documentclass{acm_proc_article-sp}

\newtheorem{theorem}{Theorem}

\newtheorem{lem}{Lemma}

\newdef{definition}{Definition}

\begin{document}

\title{Confluent Drawings: \\ Visualizing Non-planar Diagrams in a Planar Way}

\numberofauthors{2}
\author{
\alignauthor Matthew Dickerson\\ 
       \affaddr{Department of Math. and Comp. Sci.}\\
       \affaddr{Middlebury College}\\
       \affaddr{Middlebury, VT 05753}\\
       \email{dickerso@middlebury.edu}
\alignauthor \hbox{David Eppstein, Michael T. Goodrich, Jeremy Meng}
       \affaddr{Department of Info. and Comp. Sci.}\\
       \affaddr{University of California, Irvine}\\
       \affaddr{Irvine, CA 92697}\\
       \email{[eppstein, goodrich, ymeng]@ics.uci.edu}
}
\date{16 December 2002}
\maketitle
\begin{abstract}
In this paper, we introduce a new approach for drawing
diagrams that have applications in software visualization. 
Our approach is to use a technique we call
\emph{confluent drawing} for visualizing non-planar diagrams in a planar way.  
This approach allows us to draw, in a crossing-free manner, graphs---such
as software interaction diagrams---that would normally have many 
crossings.
The main idea of this approach is quite simple: we 
allow groups of edges to be merged together and drawn as ``tracks''
(similar to train tracks).
Producing such confluent diagrams automatically from a graph with
many crossings is quite challenging, however,
so we offer two heuristic algorithms to test if a non-planar graph 
can be drawn efficiently in a confluent way.
In addition, we identify several large classes of graphs that can be
completely categorized as being either
confluently drawable or confluently non-drawable.
\end{abstract}




\section{Introduction}
Software visualization is often done through the use of diagrams
constructed so that
important components, entities, agents, or objects are drawn as simple
shapes, such as circles or boxes, and relationships are drawn as
individual curves connecting pairs of these shapes.
That is, such visualizations are done by 
drawing graphs in a standard way, 
so as to assign vertices to points (or simple shapes) and to assign
edges to simple paths connecting
pairs of vertices (e.g., see~\cite{dett-gd-99,em-gda-99,kw-dgmm-01}).
Examples of such software visualizations 
include data flow diagrams~\cite{bnt-ladfd-86},
object-oriented class hierarchies~\cite{bhhkprt-tfuml-97,rjb-umlrm-98},
object-interaction diagrams~\cite{b-ooada-94},
method-call graphs~\cite{gddc-cgcoo-97,hmc-vloos-96,tp-spbcg-00},
as well as the classic application of flowcharts~\cite{k-cdf-63}
(see also~\cite{be-svl-96,pbs-ptsv-93,rc-tpvs-93,sk-mbasv-93}).
Moreover, these examples include both directed and undirected
diagrams.

In addition, it is quite common for software visualizations to be
constructed automatically rather than being hand-crafted.
Thus, there is a need for efficient algorithms that produce 
aesthetically-pleasing diagrams for software visualizations.
%
%

\subsection{Related Prior Work}
There are several aesthetic criteria that have been explored
algorithmically in the area of 
graph drawing (e.g., see~\cite{dett-gd-99,em-gda-99,kw-dgmm-01}).
Examples of aesthetic goals designed to facilitate readability
include minimizing edge crossings,
minimizing a drawing's area,
and achieving good separation of vertices, edges, and angles.
Of all of these criteria, however, the arguably most important
is to minimize edge crossings, since crossing edges tend to confuse the eye
when one is viewing adjacency relationships.
Indeed, an experimental analysis by Purchase~\cite{ref:Purchase:1997a}
suggests that edge-crossing 
minimization~\cite{jlmo-pamlc-97,jm-2lscm-97,m-amcmh-97}
is the most important
aesthetic criteria for visualizing graphs.
Ideally, we would like drawings that have no edge crossings at all.

Graphs that can be drawn in the standard way in
the plane without edge crossings are called 
\emph{planar graphs}~\cite{nc-pgta-88}, and there are a number of
existing efficient algorithms for producing crossing-free drawings of
planar graphs (e.g., see~\cite{cnao-laepg-85,con-dpgn-85,cyn-lacdp-84,%
cdgk-dpgca-99,cp-ltadp-90,dlt-pepg-84,fpp-hdpgg-90,%
gw-afdpg-00,gm-ppdga-98,k-dpguc-96,s-epgg-90,tt-pgelt-89}).

Unfortunately, most graphs are not planar; hence, most graphs
cannot be drawn in
the standard way without introducing edge crossings, and such
non-planar graphs seem to be common in software visualization
applications.
There are some heuristic algorithms for minimizing edge crossings of
non-planar graphs
(e.g., see~\cite{jlmo-pamlc-97,jm-2lscm-97,m-amcmh-97,mz-ccmp-99}),
but the general problem of drawing a non-planar graph
in a standard way that minimizes edge-crossings is NP-hard~\cite{gj-cninc-83}.
Thus, we cannot expect an efficient 
algorithm for drawing non-planar graphs so as to minimize edge crossings.

\subsection{Our Results}
Given the difficulty of edge-crossing minimization
and the ubiquity of non-planar graphs,
we explore in this paper a diagram visualization
approach, called \emph{confluent drawing},
that attempts to achieve the best of both worlds---it
draws non-planar graphs in a planar way.
Moreover, we provide two heuristic algorithms for producing confluent
drawings for directed and undirected graphs, respectively, focusing
on graphs that tend to arise in software visualizations.

\begin{figure}[htb]
  \begin{center}
  \includegraphics[width=2.75in]{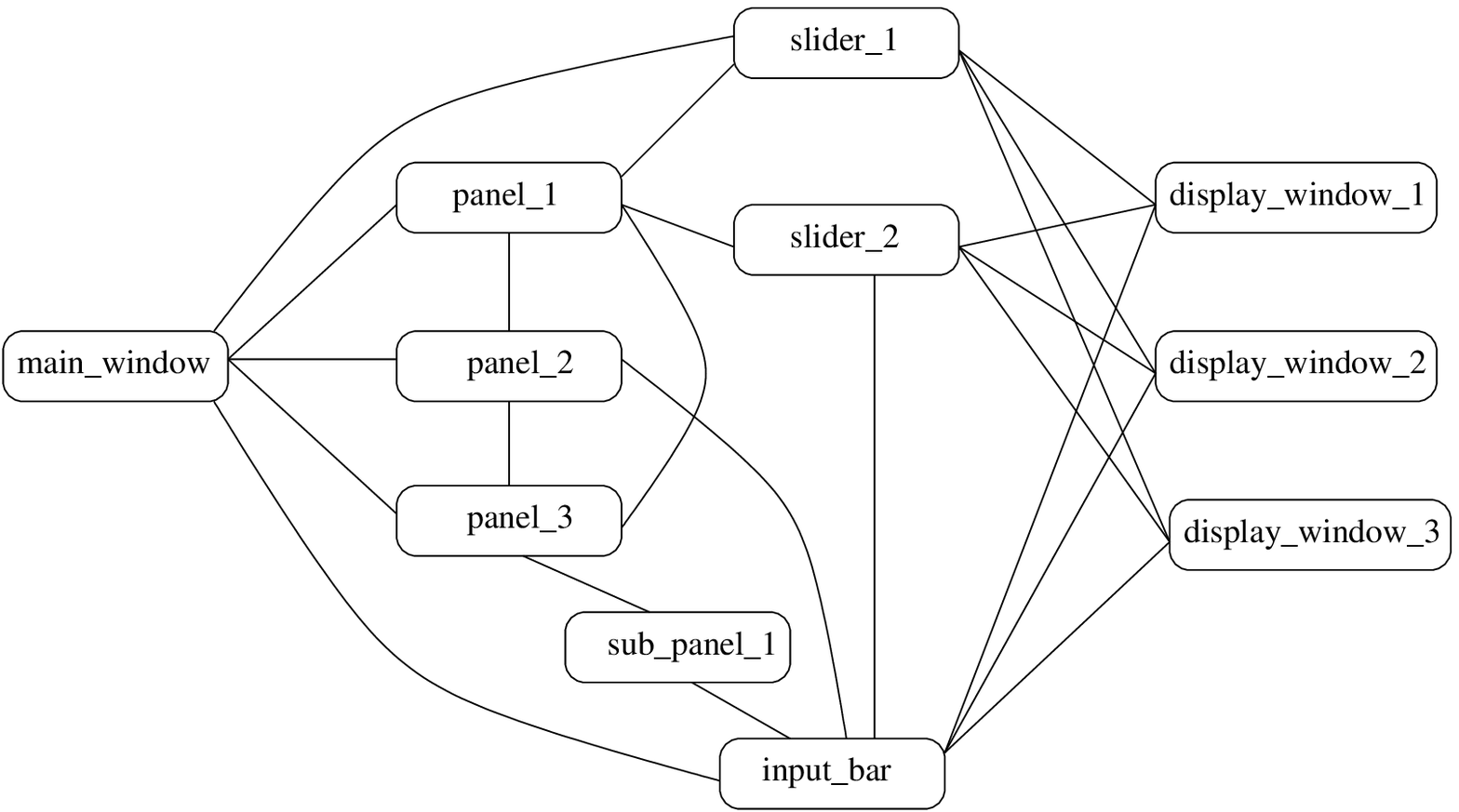} \\
  (a) \\[6pt]
  \includegraphics[width=2.75in]{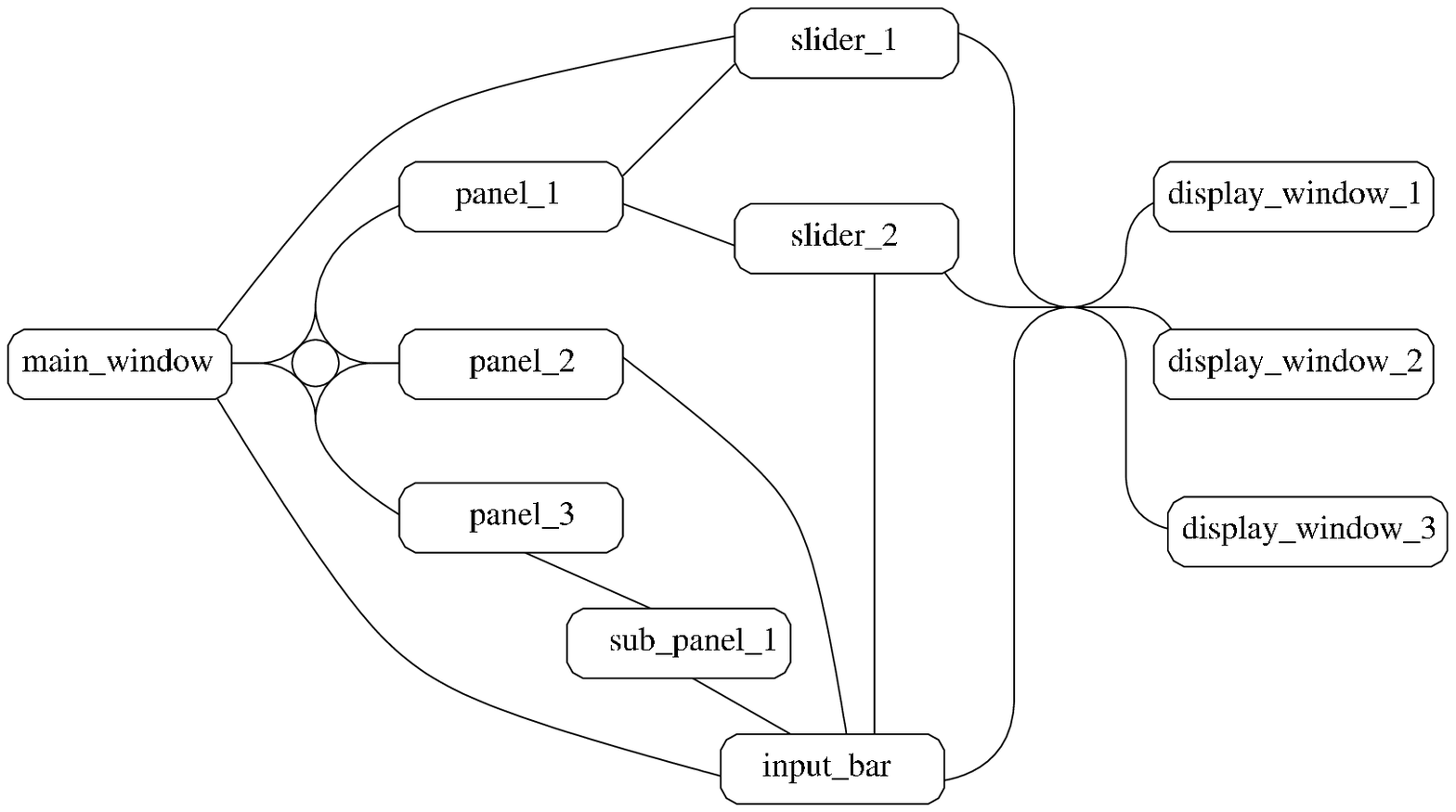} \\
  (b)
  \end{center}
  \vspace*{-10pt}
  \caption{An example of confluent drawing of an object-interaction diagram.
    Nodes here denote components in a GUI program and
    edges indicate that the adjacent
    components send messages to each other.
    We show a standard drawing in (a) and a confluent drawing in (b).
    }
  \label{fig:fig-example}
\end{figure}

The main idea of the confluent drawing
approach for visualizing non-planar graphs in a planar way
is quite simple---we merge edges into ``tracks'' so as to turn
edge crossings into overlapping paths.
(See Figure~\ref{fig:fig-example}.)
The resulting graphs are easy to read and comprehend, while also
encapsulating a high degree of connectivity information.
Although we are not familiar with any prior work on the automatic
display of graphs using this confluent diagram approach,
we have observed that some airlines use hand-crafted
confluent diagrams to display their route maps.
Diagrams similar to our confluent drawings have also been used
by Penner and Harer~\cite{ph-ctt-92} to study the topology of surfaces.

In addition to providing heuristic algorithms for recognizing and
drawing confluent diagrams, we also show that there are large classes
of non-planar graphs that can be drawn in a planar way using our
confluent diagram approach.  For example, any interval graph
or the complement of any tree can be visualized with a (planar)
confluent diagram.
Even so, we also show that there are unfortunately some graphs
that cannot be drawn in a confluent way,
including 4-dimensional hypercubes and a certain subgraph of the
Petersen graph.

This paper is organized as follows.
We give a formal definition of directed and undirected confluent
diagrams in Section~\ref{sec:confluentdrawing}.  
We describe heuristic algorithms for recognizing and drawing directed
and undirected confluent diagrams in Section~\ref{sec:heuristicalg}.
We show several special
classes of confluently drawable graphs in Section~\ref{sec:confgraphs}, 
and in Section~\ref{sec:nonconfgraphs} we demonstrate
several classes of graphs that cannot be drawn in a confluent way.


\section{Confluent Drawings}
\label{sec:confluentdrawing}

\begin{figure}[h]
\begin{center}
\includegraphics[height=1in]{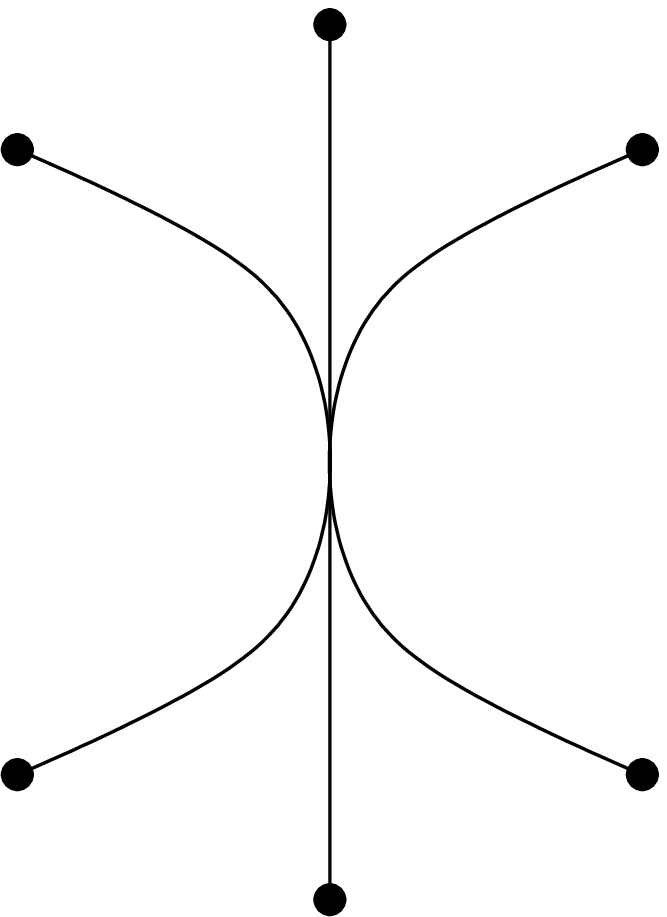}
\hspace{.5in}
\includegraphics[height=1in]{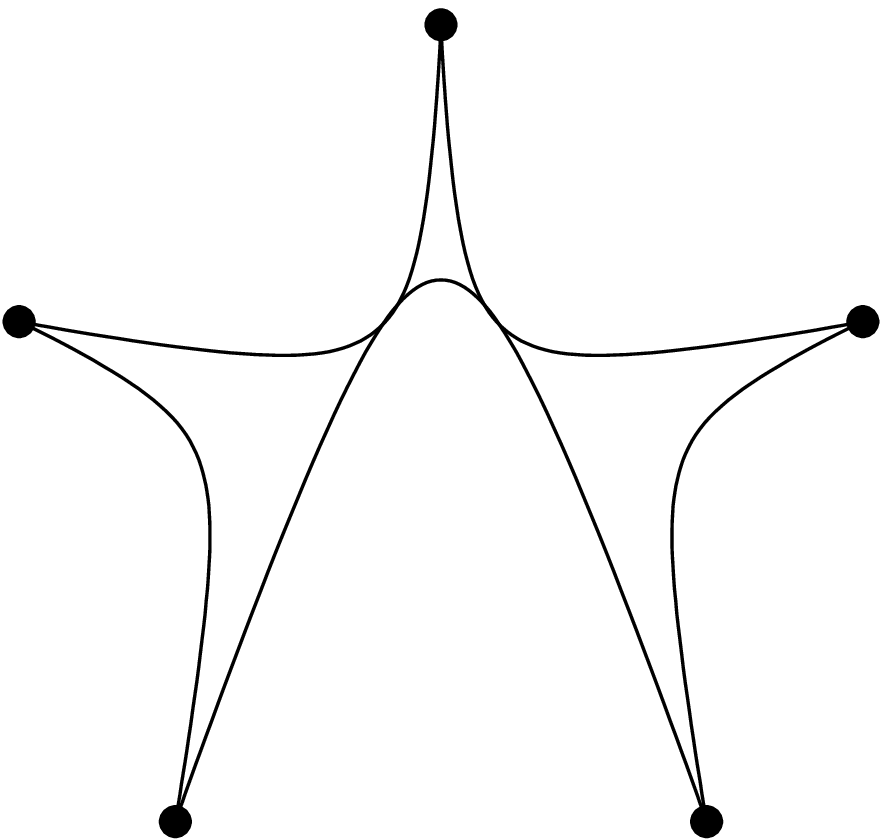}
\caption{Confluent drawings of $K_{3,3}$ and $K_5$.} \label{fig:conk33k5}
\end{center}
\end{figure}

It is well-known that every non-planar graph contains a
subgraph homeomorphic to the complete graph on five vertices, $K_5$,
or the complete bipartite graph between two sets of three vertices,
$K_{3,3}$ (e.g., see~\cite{bm-gta-76,g-agt-85}).  
On the other hand, confluent drawings, 
with their ability to merge crossing edges into single tracks,
can easily draw any $K_{n,m}$ or $K_n$ in a planar way.
Figure.~\ref{fig:conk33k5} shows
confluent drawings of $K_{3,3}$ and $K_5$.

A curve is \emph{locally-monotone} if it contains no
self intersections and no
sharp turns, that is, it contains no 
point with left and right tangents
that form an angle less than or equal to $90$ degrees.
Intuitively, a locally-monotone curve is like a single train track, which
can make no sharp turns.
Confluent drawings are
a way to draw graphs in a planar manner by
merging edges together into \emph{tracks}, which are the unions of 
locally-monotone curves.

An undirected graph $G$ is \textit{confluent} if and only if there exists a
drawing $A$ such that:
\begin{itemize}
\item 
There is a one-to-one mapping between the vertices in $G$ and
  $A$, so that, for each vertex $v \in V(G)$, there is a corresponding vertex 
  $v' \in A$, which has a unique point placement in the plane.
\item 
  There is an edge $(v_i,v_j)$ in $E(G)$
  if and only if there is a locally-monotone curve $e'$
  connecting $v_i'$ and $v_j'$ in $A$.  
\item 
$A$ is planar.
That is, while locally-monotone curves in $A$ can share overlapping portions,
no two can cross.
\end{itemize}
Our definition does not allow for confluent graphs to contain self
loops or parallel edges, although we do allow for tracks to contain
cycles and even multiple ways of realizing the same edge.
Moreover, our definition implies that
tracks in a confluent drawing have a ``diode'' property that
does not allow one to double-back or 
make sharp turns after one has started going
along a track in a certain direction. 

\begin{figure*}[t]
  \centering
  \includegraphics[width=4.4in]{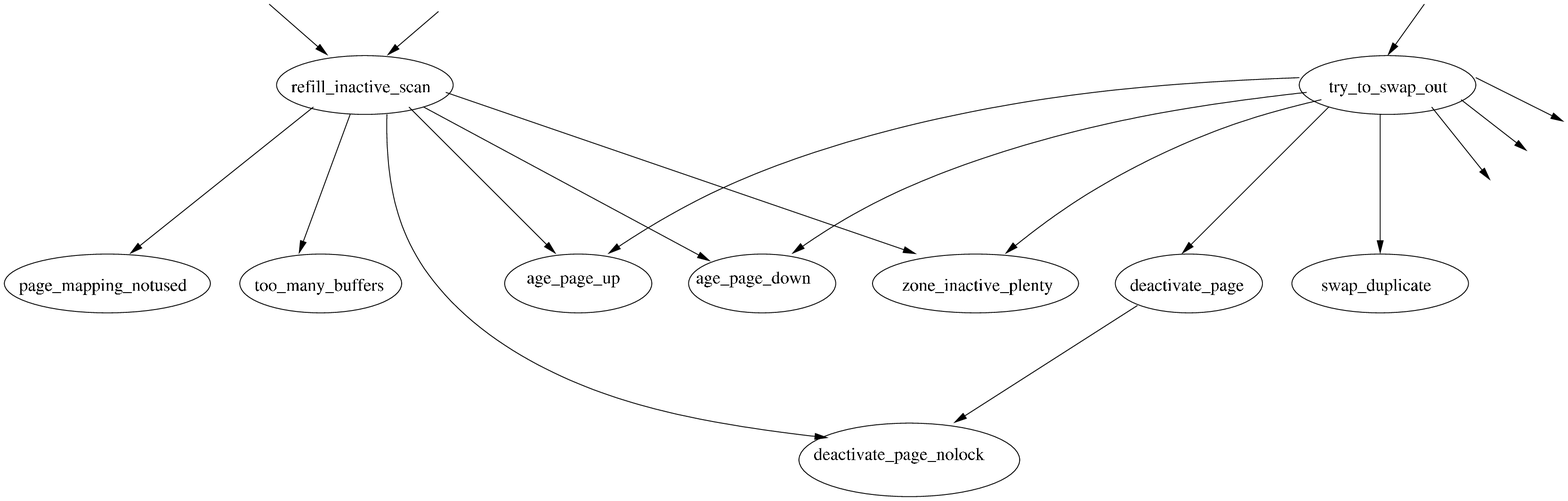}\\
  (a) \\
  \includegraphics[width=4.4in]{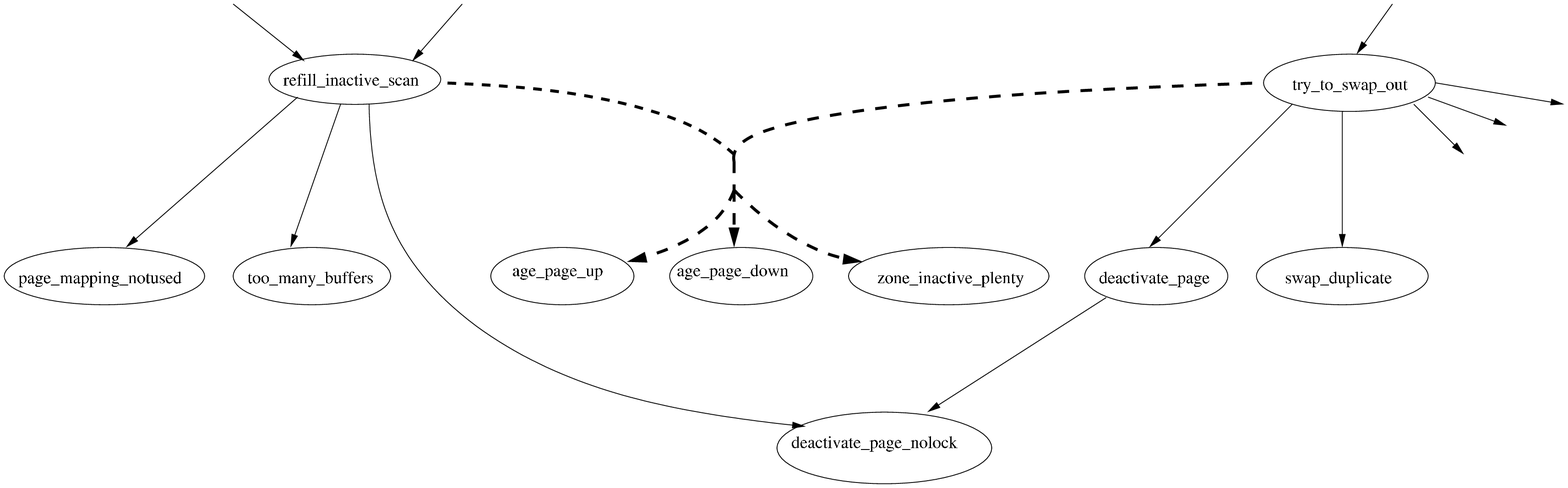} \\
  (b)
  \caption{A call graph (a) and its confluent drawing (b), 
  with the dashed part showing the confluence.}
  \label{fig:mm-call}
\end{figure*}

Directed confluent drawings are defined similarly, except that
in such drawings the locally-monotone curves are directed and the
tracks formed by unions curves must be oriented consistently.
Formally,
a directed graph $D$ is \textit{confluent} if and only if there exists
a drawing $B$ such that
\begin{itemize}
\item 
There is a one-to-one mapping between the vertices in $D$ and
  $B$, so that, for each vertex $v \in V(D)$, there is a corresponding vertex 
  $v' \in B$, which has a unique point placement in the plane.
\item 
  There is an edge $(v_i,v_j) \in E(D)$ if and only if
  there is a locally-monotone curve $e'$
  connecting $v_i'$ and $v_j'$ in $B$.  
\item 
Locally-monotone 
curves in $B$ may share some overlapping
portions, but the edges sharing the same portion of a track
must all have the same direction along that portion.
\item 
$B$ is directed and planar.
\end{itemize}

Figure~\ref{fig:mm-call} shows a part of the call graph 
of a Linux memory management module~\cite{url1-mmfig} 
and its corresponding confluent drawing. 
We choose this non-planar drawing to illustrate how confluent drawing
works, and the level information of the drawing is still preserved.
In the bottom figure we can easily tell the 
three functions (\texttt{age\_page\_up}, \texttt{age\_page\_down}, and 
\texttt{zone\_inactive\_plenty}) have two common callers \\
(\texttt{refill\_inactive\_scan} and \texttt{try\_to\_swap\_out}), while in the
original graph, it is a little more difficult to explore that information. 
One can imagine that confluent drawings can make complicated graphs 
more readable.

%
%
%
%

Confluent drawings remove crossings present in non-planar graphs,
making the graphs' structure easier to be understand.  
We feel that such drawings
may also be helpful in discovering 
certain characteristic of the graphs.
For example, given a confluent drawing,
we can easily find the 
common source vertices and destination vertices of merged edges.  
Such common structures could indicate in a method-call diagram, say, 
separate methods that can be joined together for the sake of
efficiency.
Likewise,
structures in which many sources all communicate 
with many destinations could
indicate a need for refactoring or 
lead to other useful insights about a software design.


\pagebreak
\section{Heuristic Algorithms} 
\label{sec:heuristicalg}

Though the planarity of a graph can be tested in linear time,
it appears difficult to quickly 
determine whether or not a graph can be drawn confluently.  If 
a graph $G$ contains a non-planar subgraph, then $G$ itself is non-planar 
too.  But similar closure properties are not true for confluent graphs.
Adding vertices and edges to a non-confluent graph
increases the chances of edges 
crossing each other, but it also increases the chances of edges merging.
Currently, the best method we know of for determining conclusively
in the worst case whether a graph is
confluent or not is a brute force one of
exhaustively listing all possible ways of edge merging and checking the
merged graphs for planarity.  Therefore, it is of interest to develop
heuristics that can find confluent drawings in many cases.

\begin{figure}[htb]
  \centering
  \includegraphics[scale=.6]{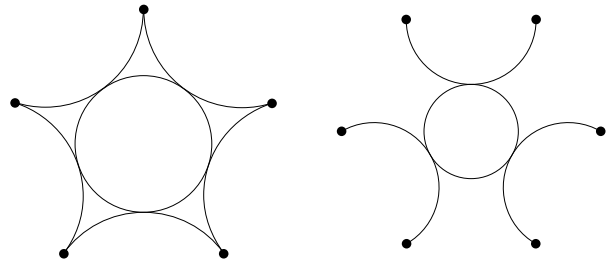}
  \caption{Confluent drawings of $K_5$ and $K_{3,3}$ using ``traffic
  circle'' structures.}
  \label{fig:traffic-cir}
\end{figure}

Figure~\ref{fig:traffic-cir} shows confluent drawings using a ``traffic circle'' structure
for complete graphs and
complete bipartite subgraphs.
At a high level, our heuristic drawing algorithm
iteratively 
finds clique subgraphs and biclique subgraphs and replaces them with 
traffic-circle subdrawings.

Chiba and Nishizeki~\cite{cn-asla-85} discuss the problem of
listing complete subgraphs (cliques) for graphs of bounded
arboricity.
The \textit{arboricity} $a(G)$ is the minimum number of forests into 
which the edges of $G$ can be partitioned.  
A bounded arboricity is equivalent to a notion of sparsity.  We 
believe graphs arising in software visualization are often
likely to be sparse, thus 
the listing algorithm is applicable for such graphs.
Chiba and Nishizeki show 
that there can be at most $O(n)$ cliques of a given size in such
graphs and give a linear time algorithm for listing these clique
subgraphs.
Eppstein \cite{Epp-IPL-94} gives a linear time algorithm for 
listing maximal complete bipartite subgraphs (bicliques) in  
graphs of bounded arboricity.  The total complexity of all such graphs is $O(n)$,
and again they can be listed in linear time.

In our heuristic algorithm for undirected graphs, we will use the
clique subgraphs listing 
and the biclique subgraphs listing algorithms as our subroutines.

\noindent\textsc{HeuristicDrawUndirected}($G$)\\
\textit{Input.} A undirected sparse graph $G$.\\
\textit{Output.} Confluent drawing of $G$ if succeed, fail otherwise.\\
1.  If $G$ is planar\\
2.  \quad draw $G$\\
3.  else if $G$ contains a large clique or biclique subgraph $C$\\
4.  \quad create a new vertex $v$\\
5.  \quad obtain a new graph $G'$ by removing edges of $C$ 
and connecting each vertex of $C$ to $v$\\
6.  \quad \textsc{HeuristicDrawUndirected($G'$)} \\
7.  \quad replace $v$ by a small ``traffic circle'' to get a
confluent drawing of $G$ \\
8.  else fail\\

In step 3,  the cliques are given higher priority over bicliques,
otherwise a clique would be partially covered by a biclique.
Cliques of three or fewer vertices, and bicliques with one side consisting of only one vertex,
are not replaced because the replacement cannot change the planarity of the graph.
We now discuss the time
performance of this heuristic.

\begin{theorem}
In graphs of bounded arboricity, algorithm \textsc{HeuristicDrawUndirected}
can be made to run in time $O(n)$, assuming hash tables with constant time per operation.
\end{theorem}

\begin{proof}
We store a bit per edge of the original graph so we can quickly look up
whether it is still part of our replacement.  We begin the heuristic
by looking for cliques, since we want to give them priority over
bicliques.  List all the complete subgraphs in the graph with four or more
vertices, and sort them by size.  Then, for each complete subgraph $X$ in sorted order,
we check whether $X$ is still a clique of the modified graph, and if so perform
a replacement of $X$.  It is not hard to see that the new vertex $v$ of
the replacement cannot belong to any clique, so this algorithm correctly
finds a maximal sequence of cliques to replace.

Next, we need to similarly dynamize the search for bicliques.
This is more difficult, because a biclique may have nonconstant size and
because the replacement vertex $v$ may belong to additional bicliques.
We perform this step by dynamizing the algorithm of Eppstein~\cite{Epp-IPL-94} for listing
all bicliques.  This algorithm uses the idea of a {\em $d$-bounded acyclic orientation}:
that is, an orientation of the edges of the graph, such that the oriented graph is acyclic
and the vertices have maximum outdegree $d$.  For graphs of arboricity $a$, a
$(2a-1)$-bounded acyclic orientation may easily be found in linear time.
For such an orientation, define a {\em tuple} to be a subset of the outgoing neighbors
of any vertex, and let $v$ be a {\em tuple creator} of tuple $T$ if all vertices of $T$ are
outgoing neighbors of $v$.  For graphs of bounded arboricity, there are at
most linearly many distinct tuples.  For each maximal biclique,
one of the two sides of the bipartition must be a tuple, $T$~\cite{Epp-IPL-94}.
The other side consists of two types of vertices:
tuple creators of $T$, and outgoing neighbors of vertices of $T$.

Our algorithm stores a hash table indexed by the set of all tuples in
the modified graph. The hash table entry for tuple $T$ stores the number
of tuple creators of $T$, and a list of outgoing neighbors of vertices
of $T$ that are adjacent to all tuple members.  For each edge 
$(u,v)$ in the graph,
oriented from $u$ to $v$,
we store a list of the tuples $T$ containing $v$ for which $u$ is listed as an outgoing neighbor. 
We also store a priority
queue of the maximal bicliques generated by each tuple, prioritized by
size; it will suffice for our purposes if the time to find the largest
biclique is proportional to the biclique size, and it is easy to
implement a priority queue with such a time bound.
With these structures, we may easily look up each successive biclique replacement
to perform in algorithm \textsc{HeuristicDrawUndirected}.
Each replacement takes time proportional to the number of edges removed from the graph,
so the total time for performing replacements is linear.

It remains to show how to update these data structures when we perform a
biclique replacement. To update the acyclic orientation, orient each
edge from $C$ to $v$, except for those edges from vertices of $C$ that
have no outgoing edges in $C$.  It can be seen that this orientation
preserves $d$-boundedness and acyclicity. When a new vertex $v$ is
created by a replacement, create the appropriate hash table entries for
tuples containing $v$; the number of tuples created by a replacement is
proportional to the number of edges removed in the same replacement, so
the total number of tuples created over the course of the algorithm is
linear. Whenever a replacement causes edges from a vertex $x$ to change,
update the hash entries for all tuples for which $x$ is a creator; this
step takes $O(1)$ time per change.  Also, update the hash entries for
all tuples to which $x$ belongs, to remove vertices that are no longer
outgoing neighbors of $x$; this step takes time $O(1)$ per changed
tuple, and each tuple changes $O(1)$ times over the course of the
algorithm. Whenever a change removes incoming edges of~$x$, we must remove the other endpoints of those edges from the
lists of outgoing neighbors of tuples to which $x$ belongs; using the
lists associated with each incoming edge, this takes constant time per
removal. Therefore, all steps can be performed in linear total time.
\end{proof}

\begin{figure}[htb]
  \centering
  \includegraphics[scale=.5]{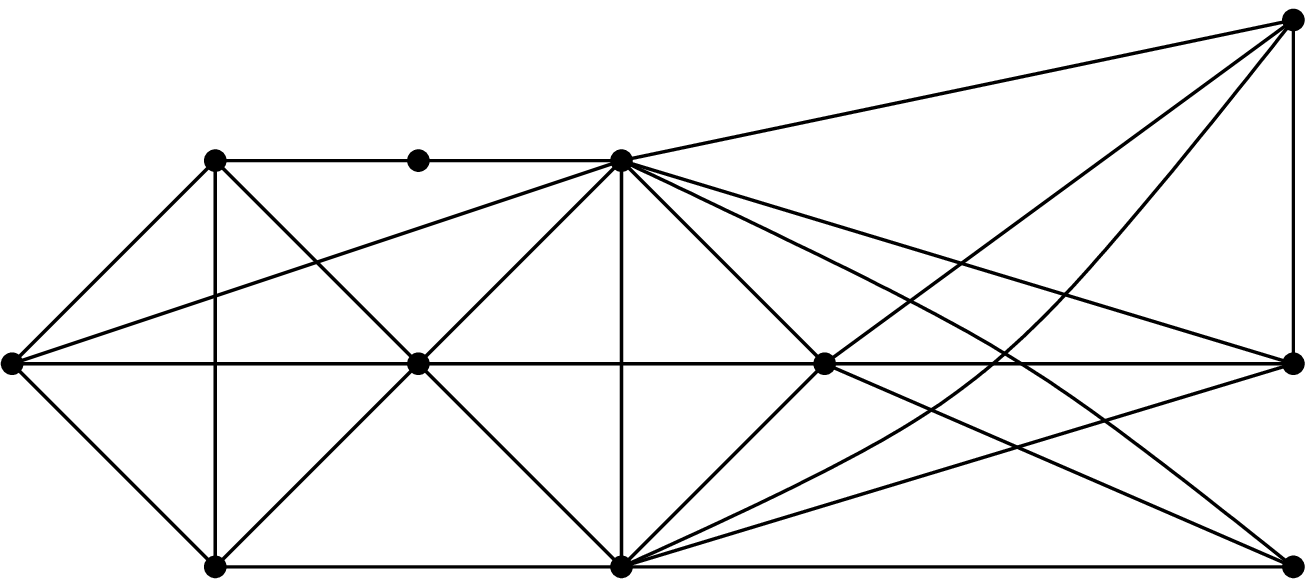} \\
  (a) \\[6pt]
  \includegraphics[scale=.5]{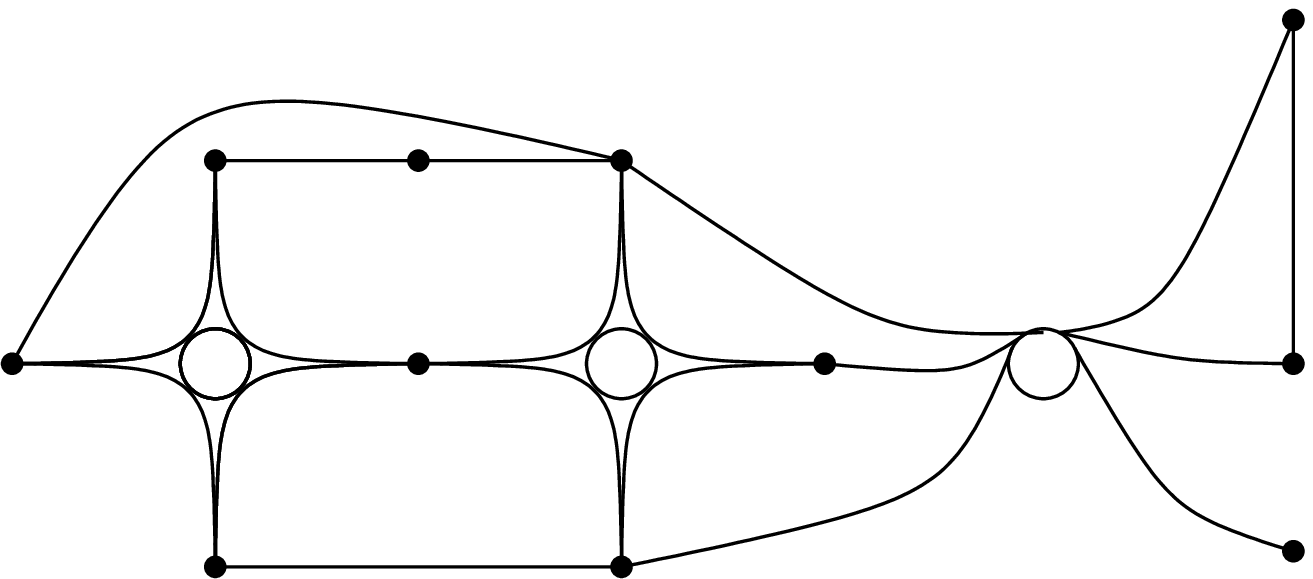} \\
  (b)
  \vspace*{-10pt}
  \caption{An example of running the undirected heuristic
    algorithm. The input graph is shown in (a) and the output drawing
    is shown in (b).}  
  \label{fig:example}
\end{figure}

An example of the input for algorithm \textsc{Heuris\-tic\-Draw\-Un\-di\-rected}
and the output drawing produced by this heuristic is shown in
Figure~\ref{fig:example}. 

\begin{figure}[hbt]
  \centering
  \includegraphics[scale=.4]{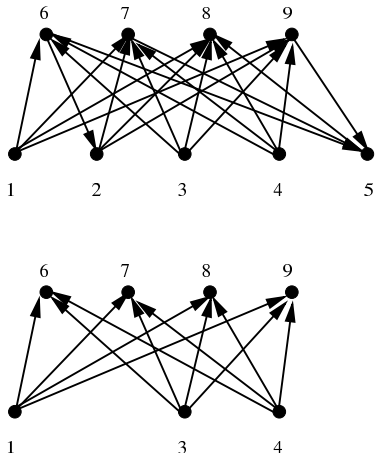}
  \caption{Maximal directed complete bipartite subgraphs.}
  \label{fig:di-mcbs}
\end{figure}

For directed graph, the algorithm is slightly different. Because 
the tracks in directed confluent drawings are required to have  
directions, the ``traffic circle'' structure will not work for
directed cliques.
Thus we only look for directed bicliques in step 3 in the directed
version of the heuristic algorithm.  
Next we discuss how to find maximal directed bicliques.
Maximal directed complete bipartite subgraphs in a sparse directed graph 
$G$ can be found by first listing maximal undirected complete bipartite 
subgraphs in the underlying undirected graph of $G$.  Then for each of 
these subgraphs examine the corresponding directed subgraph.  We choose 
the side of the bipartition with larger size and partition it according to 
how their edges are oriented to the other side of the bipartition 
(In Figure~\ref{fig:di-mcbs}, the bottom directed $K_{3,4}$ is obtained 
from the top graph).  

\section{Some Confluent Graphs}
\label{sec:confgraphs}

The heuristic algorithms 
presented in the previous section are most applicable to
sparse graphs, because sparseness is needed for the linear time bound
of the maximal bipartite subgraph listing subroutine.
However, there are also several denser classes of graphs that we can
show to be confluent.

\pagebreak
\subsection{Interval graphs}
An \emph{interval graph} is formed by a set of closed intervals
$S=\{[a_1,b_1],[a_2,b_2],\ldots,[a_n,b_n]\}$.
The interval graph is defined to have the 
intervals in $S$ as its vertices and two vertices $[a_i,b_i]$ and
$[a_j,b_j]$ are connected by an edge if and only if these two
inverals have a non-empty intersection.
Such graphs are typically non-planar, but we can draw them in a
planar way using a confluent drawing\footnote{%
  A similar construction works for circular-arc graphs and is left as
  an exercise for the interested reader.
}.

\begin{theorem}
Every interval graph is confluent.
\end{theorem}

\begin{figure}[htb]
\begin{center}
\includegraphics[scale=.7]{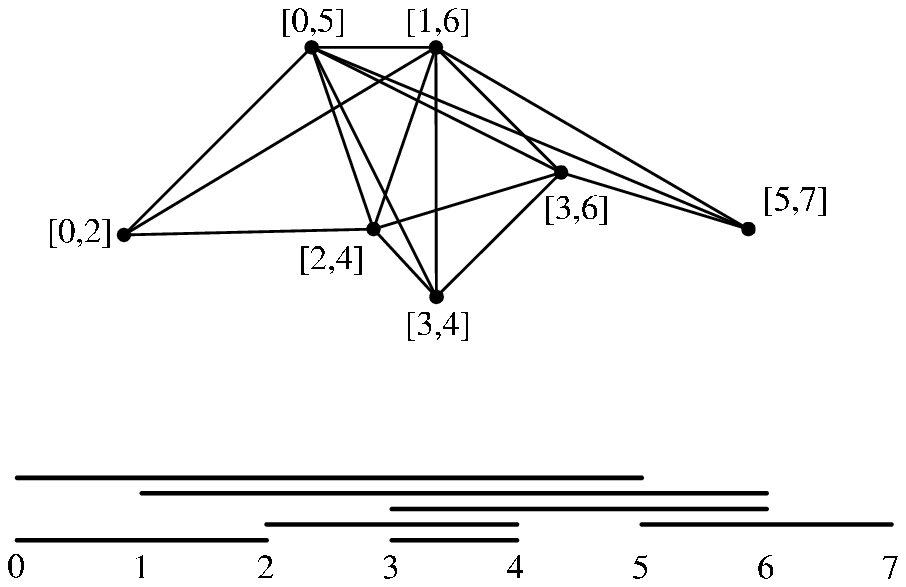} \\
(a) \\[6pt]
\includegraphics[scale=.7]{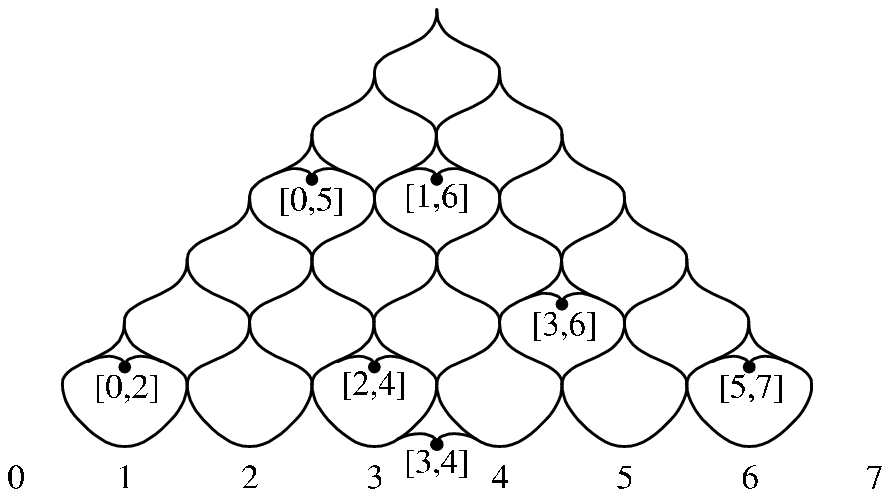} \\
(b)
\caption{Illustrating a confluent way to draw a non-planar interval graph:
(a) an interval garph and its defining intervals;
(b) its corresponding confluent drawing.}
\label{fig:interval}
\end{center}
\end{figure}

\begin{proof}
The proof is by construction.
We number the interval endpoints by rank, $X=\{0,1,\ldots,n-1\}$, and place
these endpoints along the $x$-axis.
We then build a two-dimensional lattice 
on top of these points in a fashion similar to Pascal's triangle,
using a connector similar to an upside-down ``V''.
These connectors stack on top of one another so that the apex
of each is associated with a unique interval on $X$.
We place each point from our set $S$ of intervals just under its
corresponding apex and connect it into the (single) track so that it
can reach everything directly dominated by this apex in the lattice.  
At the bottom level, we connect the updside-down
V's with rounded connectors.  By this contruction, we create a single
track that allows each pair of vertices connected in the interval graph to have
a locally-monotone path connecting them.
(See Figure~\ref{fig:interval}.)
\qed
\end{proof}

\subsection{Complements of trees}
\label{sec:cotree}

The complements of trees (graphs formed by connecting all pairs of
vertices that are not connected in some tree) are also called
\textit{cotrees}.  In general, cotrees are highly non-planar and
dense, since a cotree with $n$ vertices has $n(n-1)/2-n+1$ edges.
Nevertheless, we have the following interesting fact.

\begin{theorem}
The complement of a tree is confluent.
\end{theorem}

\begin{figure}[hbt!]
\centerline{\includegraphics[scale=.42]{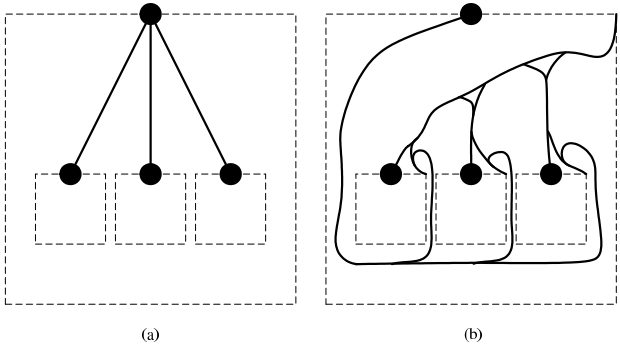}}
\caption{Illustrating a confluent way to draw the complement of a tree:
(a) a node and its children in the tree; 
(b) the corresponding portion of a track in
the confluent drawing of the complement.}
\label{fig:tree-neg}
\end{figure}

\begin{proof}
We prove the claim by recursive construction, using a single track for the
entire graph.
Assign a bounding rectangle for the tree and a bounding rectangle for every
subtree in that tree. Place the complement of the tree into the bounding
rectangles such that nodes of every subtree is within its bounding rectangle
and the bounding rectangles of subtrees are contained in their parent's
bounding rectangle. 
In addition, place a connector at the Northeastern corner of every bounding
box. This connector is an imaginary point at which
the single track for the entire graph will connect into this portion
of the cotree.
(See Figure~\ref{fig:tree-neg}.) Connect the root node
in each subtree to every connector of its children. Connect every node to the
connector of its parent. Also connect every node to its siblings and the
connectors of its siblings, as shown in the figure. 
The obtained drawing is the confluent drawing of
the complement of the given tree.
\qed 
\end{proof}

\begin{figure}[hb]
\begin{center}
\includegraphics[scale=.3]{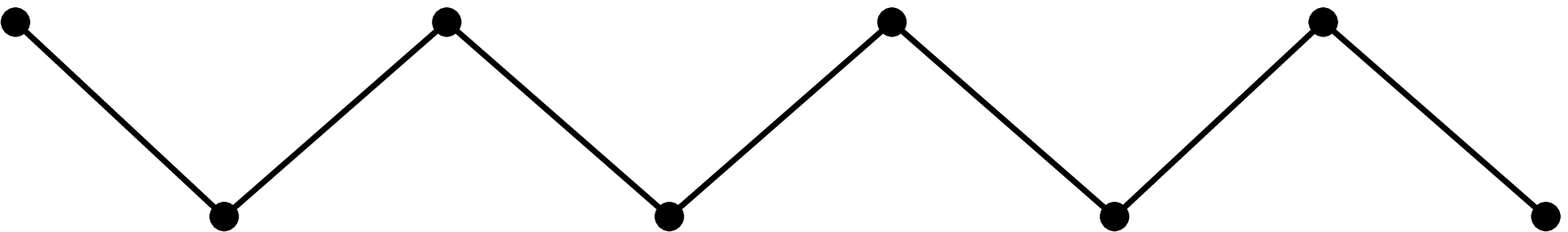}
\\[0.1in]
\includegraphics[scale=.3]{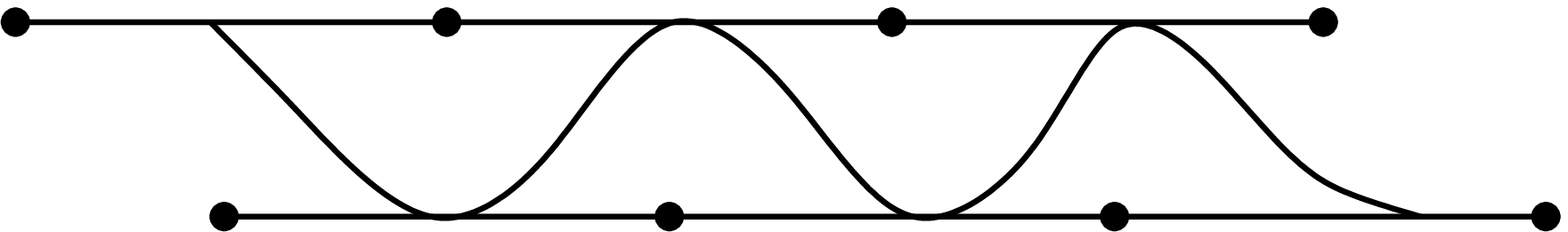}
\caption{A path and one confluent drawing of its complement.} \label{fig:path}
\end{center}
\end{figure}

Paths are very special cases of trees. Every vertex in a path has a degree of
$2$ except its two endpoints, each of which has a degree of $1$. The
complement of a path can be drawn using the cotree method in the above proof.
We show a nice confluent drawing of the complement of a path
in Figure~\ref{fig:path}.

\subsection{Cographs}
\label{sec:cographs}

A \textit{complement reducible graph} (also called a \textit{cograph}) is
defined recursively as follows \cite{ CORNEILdg}:

\vspace*{-8pt}
\begin{itemize}
\setlength{\itemsep}{0pt}
\setlength{\parsep}{0pt}
  \item A graph on a single vertex is a cograph.
  \item If $G_1$, $G_2$, $\cdots$, $G_k$ are complement reducible
  graphs, then so is their union $G_1 \cup G_2 \cup \cdots \cup G_k$.
  \item If $G$ is a complement reducible graph, then so is its
  complement $\overline{G}$.
\end{itemize}

\vspace*{-8pt}
Cographs can be obtained from single node
graphs by performing a finite number of unions and complementations.

\begin{theorem}
Cographs are confluent.
\end{theorem}

\begin{figure}[hb]
  \centering
  \includegraphics[scale=.4]{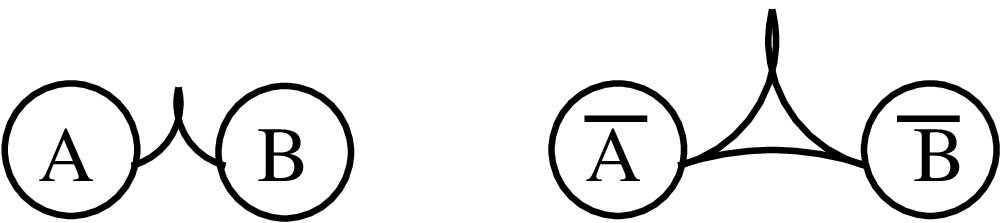}
  \vspace*{-10pt}
  \caption{Confluent $A\cup B$ and $\overline{A\cup B}$.}
  \label{fig:cogrdr}
\end{figure}

\begin{proof}
If cographs $A$ and $B$ are confluent, we can show $A \cup B$ and $\overline{A
\cup B}$ are confluent too.  First we draw $A$ confluently inside a disk and
attach a ``tail'' to the boundary of the disk.  Connect the attachment point
to each vertex in the disk. $B$ is drawn in the same way.  Then $A \cup B$ is
formed by joining the two ``tail'' together so that they don't connect to each
other. $\overline{A \cup B}$ is formed by joining the two ``tails'' of
$\overline{A}$ and $\overline{B}$ together so that they connect to each other.
(See Figure~\ref{fig:cogrdr}.)
By the definition of cographs and induction we know cographs are confluent.
\qed \end{proof}

\begin{figure}[hb]
\vspace*{-6pt}
\begin{center}
\includegraphics[scale=.5]{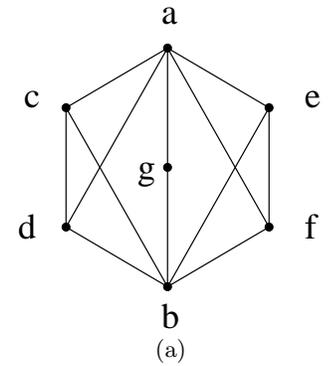} \\
(a) \\[4pt]
\includegraphics[scale=.4]{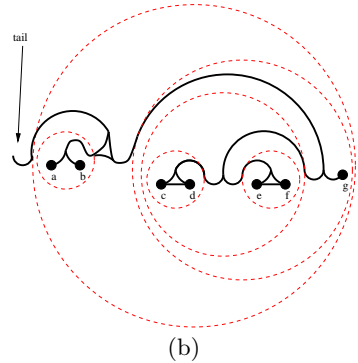} \\
(b)
\end{center}
\vspace*{-15pt}
\caption{Confluent drawing of a cograph $\overline{\cup}(\overline{\cup}(a,b),
\overline{\cup}(\overline{\cup}(c,d),\overline{\cup}(e,f),g))$. Imaginary 
disks are drawn in dashed circle.}
\label{fig:cograph}
\end{figure}

\clearpage
\subsection{Complements of $n$-cycles}
\label{sec:cocycle}

A $n$-cycle is a cycle with $n$ vertices. 

\begin{theorem}
The complement of an $n$-cycle is confluent.
\end{theorem}

\begin{proof}
First remove one vertex from the $n$-cycle and draw the confluent graph for
the complement of the obtained path.  Then add the vertex back and connect it
with all vertices in the path except for its two neighbors. The obtained
drawing is a confluent drawing.
\qed \end{proof}

\begin{figure}[hb]
\begin{center}
\includegraphics[scale=.3]{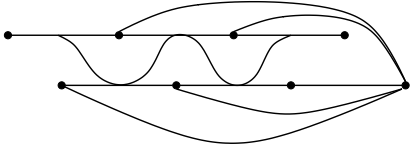}
\caption{A confluent drawing of $\overline{C}_8$.}
\label{fig:cycle-neg}
\end{center}
\end{figure}

An example of drawing a cocycle confluently is shown in
Figure~\ref{fig:cycle-neg}.

\section{Some Non-confluent Graphs}
\label{sec:nonconfgraphs}

In this section,
we show that some graphs cannot be drawn confluently.  These graphs 
include 
the Peterson graph $P$, the graph $P-v$ 
formed by removing one vertex from Peterson graph, 
graphs formed by subdividing every edge of non-planar graphs,
and
the $4$-dimensional hypecube.

\subsection{The Petersen graph}
\label{sec:Petersen}
\label{sec:smallest}

\begin{figure}[ht]
  \centering
  \includegraphics[scale=.6]{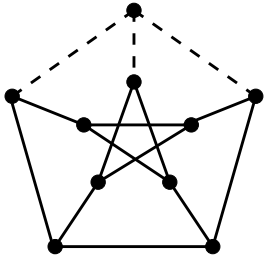}
  \caption{The Petersen graph. The edges incident on one of the vertices
  are shown dashed.}
  \label{fig:peter}
\end{figure}

\begin{figure}[ht]
  \centering
  \includegraphics[scale=.4]{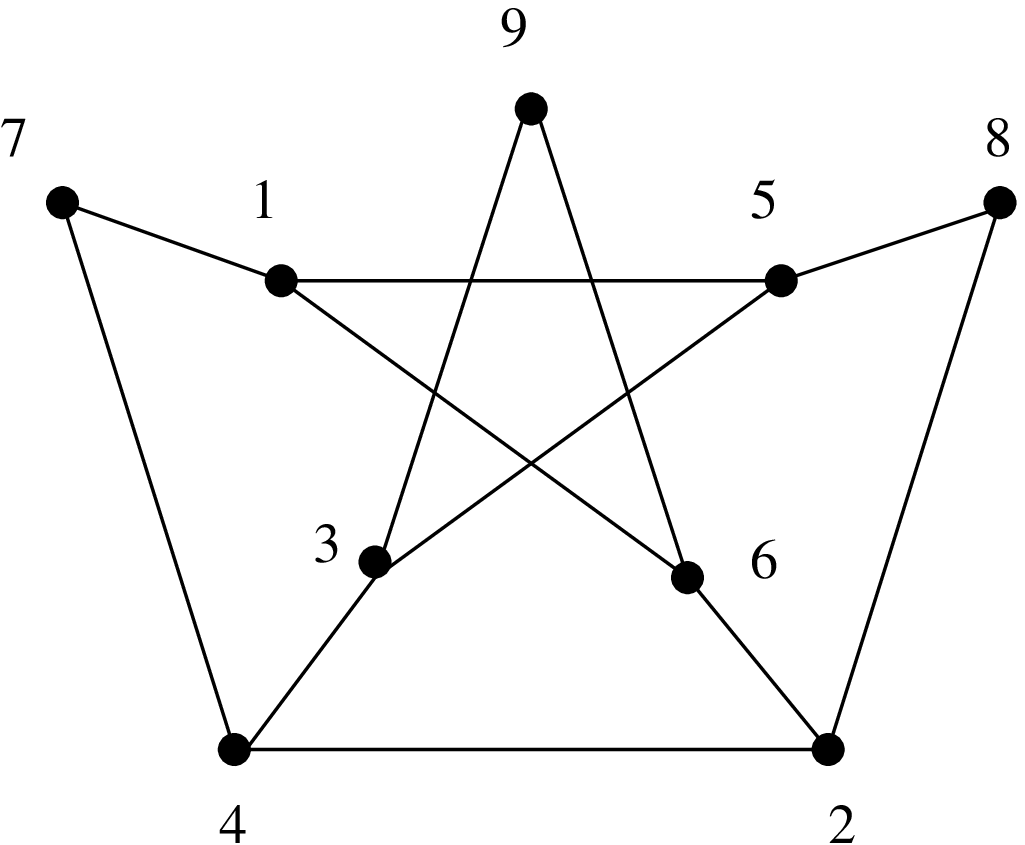}\\
  \includegraphics[scale=.4]{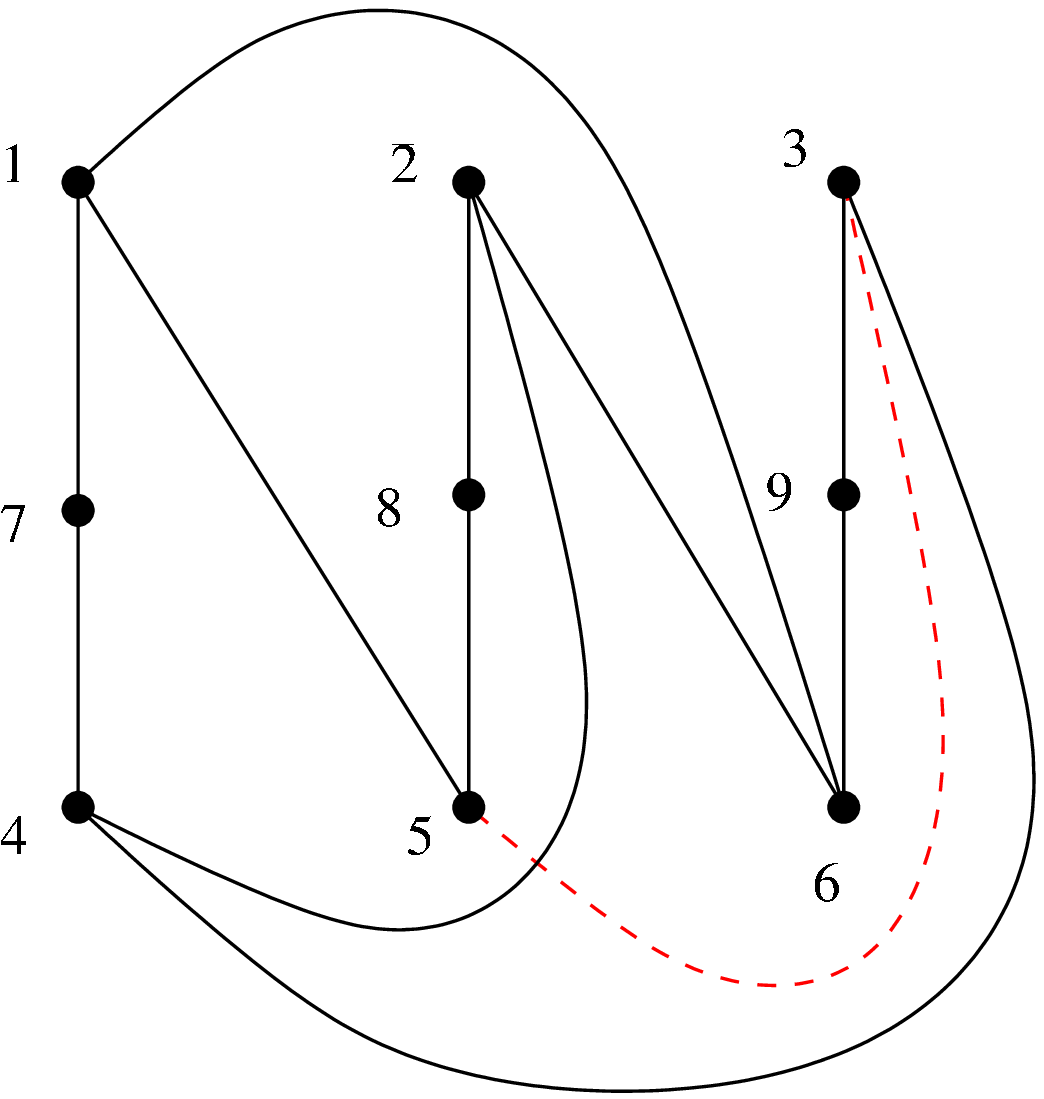}
  \caption{Removing one vertex of the Petersen graph produces a subdivision of $K_{3,3}$.}
  \label{fig:Petersen-1}
\end{figure}

By removing one vertex and its incident edges from the Petersen graph
(Figure~\ref{fig:peter})
we obtain  
a graph homeomorphic to $K_{3,3}$.  It contains no $K_{2,2}$ as a
subgraph. Moreover, note that $K_{2,2}$ is the 
most basic structure that allows for edge merging into tracks. 
Thus the resulting graph is non-confluent. 
This graph, shown in Figure~\ref{fig:Petersen-1}, is the smallest non-confluent graph we know of.

The Petersen graph itself is also non-confluent, as adding the vertex and edges back to its non-confluent subgraph doesn't create any four-cycles that could be used for confluent tracks.

\pagebreak
\subsection{Other non-confluent graphs}

If we subdivide every edge of a non-planar graph, by adding a single
vertex in the ``middle'' of each edge, the resulting graph is 
non-confluent, because the new vertices do not take part in any 4-cycles
and so can not be included in any confluent tracks.
For the same reason, if, for each edge of a non-planar graph, we add a new vertex 
and connect this new vertex to the both end points of that edge, the 
result is also non-confluent.   In particular, adding new vertices in this way to the graph $K_5$ produces a non-confluent chordal graph, so despite our proofs that other graph families with tree-like structures are confluent, chordal graphs are not all confluent.


\subsection{4-dimensional cube}
\label{sec:cube4d}

\begin{figure}[ht]
  \centering
  \includegraphics[scale=.4]{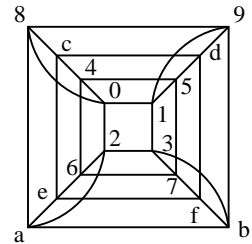} 
  \caption{The $4$-dimensional hypercube.}
  \label{fig:cube4d}
\end{figure}

The $4$-dimensional hypercube in Figure~\ref{fig:cube4d} (a) is
non-confluent. 

The hypercube contains many subgraphs isomorphic to $3$-dimensional cubes.
Cubes are planar graphs, but
in order to show non-confluence for the hypercube, we analyze more carefully
the possible drawings of the cubes.  Observe that, because 
there are no $K_{2,3}$ subgraphs in cubes or hypercubes,
the only possible confluent tracks are $K_{2,2}$'s formed from the vertices of a single cube face.

\begin{figure*}[t]
$$
{\includegraphics[scale=.4]{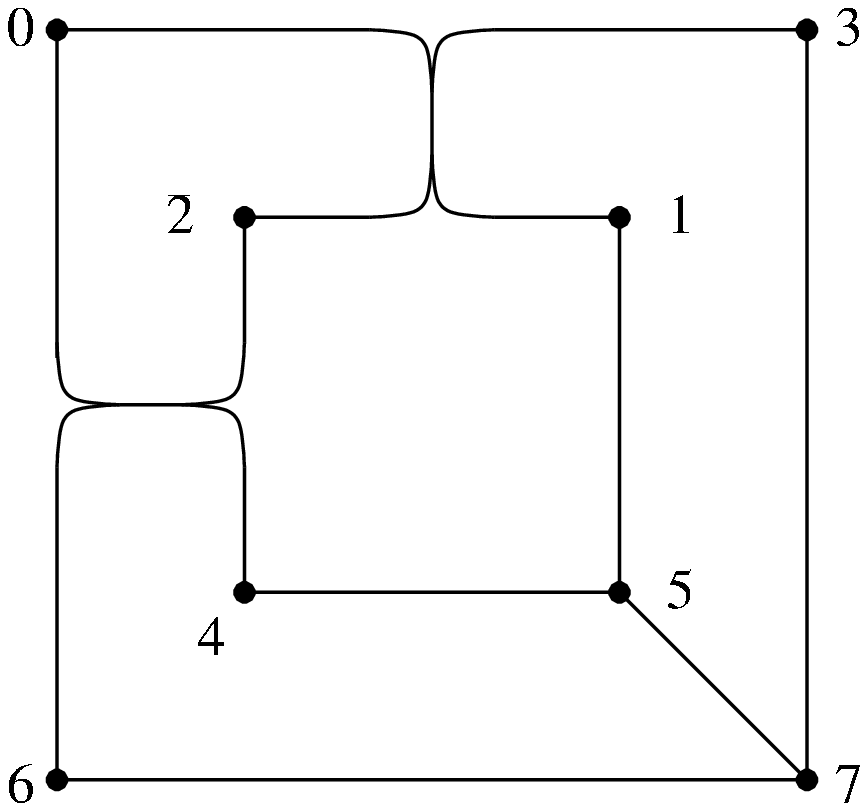}\atop\hbox{(a)}}\hbox{\hspace{10pt}}
{\includegraphics[scale=.4]{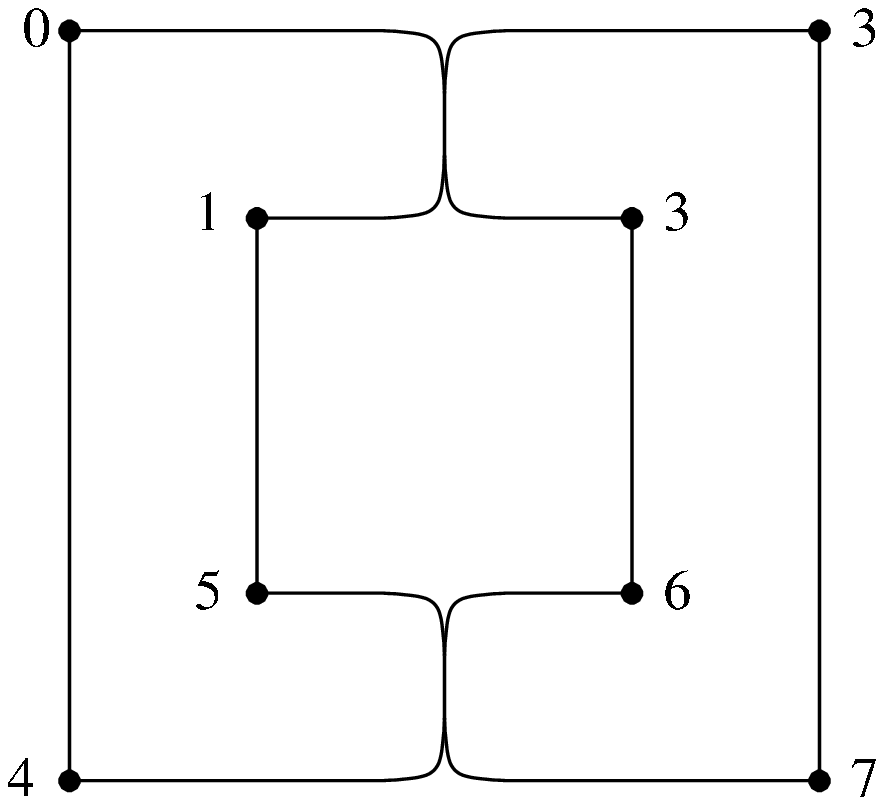}\atop\hbox{(b)}}\hbox{\hspace{10pt}}
{\includegraphics[scale=.4]{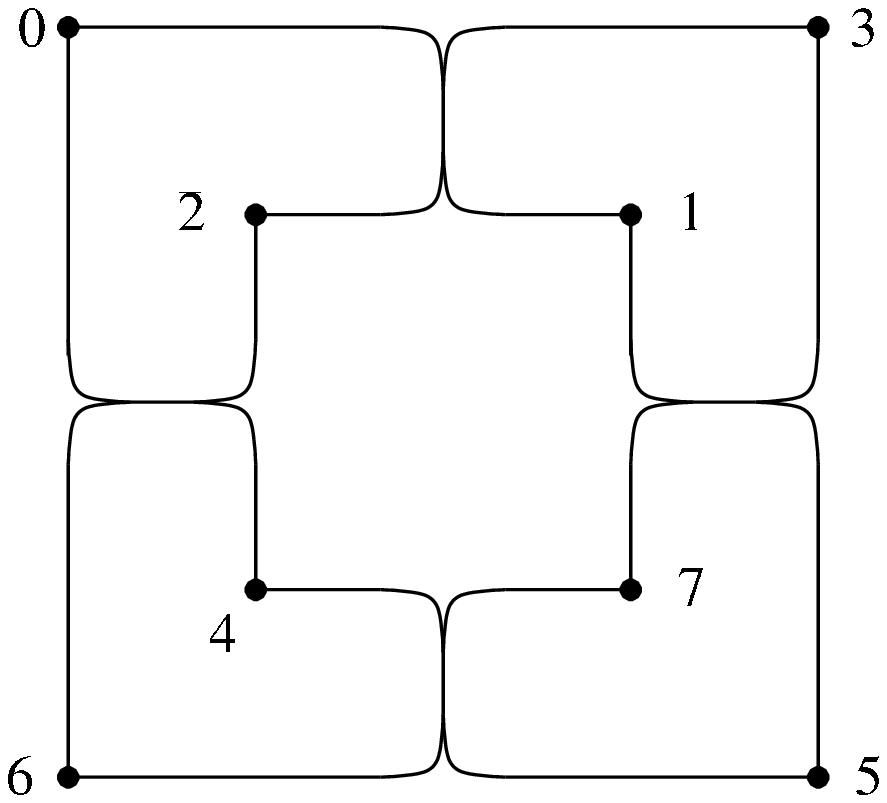}\atop\hbox{(c)}}\hbox{\hspace{10pt}}
$$
  \caption{Three confluent drawings of a cube.}
  \label{fig:cubedrawings}
\end{figure*}

\begin{lem}
A cube has only the four confluent drawings shown in 
Figure~\ref{fig:cubedrawings},
or combinatorially equivalent rearrangements of these drawings 
in which we choose a different face as the outer one.
\end{lem}

\begin{proof}
For convenience, we
consider the drawings to be on a sphere instead of
in the plane, so the outer face is not distinguished. Every cube face
can be drawn either as a quadrilateral or as a track in a confluent
drawing of $K_{2,2}$.  
We divide into cases based on the number of cube faces replaced by
tracks. 

{\bf Case 0:} No faces are replaced by tracks.  We get the usual
planar drawing of a cube. It is unique because a cube is
$3$-connected.

{\bf Case 1:} One face is replaced by a track.  This case is not
possible, because the underlying graph of the drawing (formed by
placing new vertices at track junctions) is non-planar.

{\bf Case 2:} Only two adjacent faces are replaced by tracks.  We have
the drawing of Figure~\ref{fig:cubedrawings} (a).  It is unique
because the underlying planar graph is $3$-connected.

{\bf Case 3:} Two opposite faces are replaced by tracks.  We have
the drawing of Figure~\ref{fig:cubedrawings} (b).  It is unique
because the underlying planar graph is $3$-connected.

{\bf Case 4:} Three mutually adjacent faces are replaced by
tracks. This case is not possible, even if we allow additional faces to be replaced
by tracks as well.  For, suppose the faces 
$0-1-3-2$,
$0-2-6-4$, and $0-1-5-4$ are replaced by tracks. The underlying graph
of these replaced edges has a drawing with four faces, in which
vertices $3$, $5$, and $6$ are dangling and may each go in either of
two faces (Figure~\ref{fig:cubetriple}).  However, it is not
possible for all three to be in the same face.  So they can't all
three be connected to vertex $7$, as edges to $7$ can not cross the existing
tracks. 

\begin{figure}[ht]
  \centering
  \includegraphics[scale=.4]{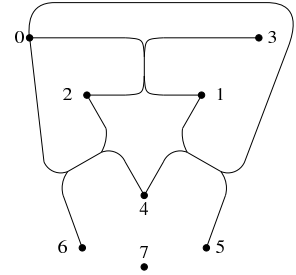}
  \caption{Attempt to use confluent tracks for three mutually-adjacent faces of a cube.}
  \label{fig:cubetriple}
\end{figure}

{\bf Case 5:} Three non-mutually adjacent faces are replaced by
tracks. This case is not possible because the underlying graph is
non-planar (Figure~\ref{fig:threenonadj}). 

\begin{figure}[ht]
  \centering
  \includegraphics[scale=.4]{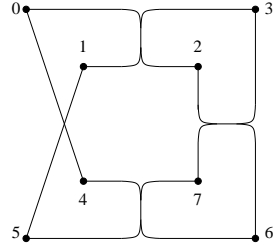}
  \caption{Attempt to use confluent tracks for three non-mutually-adjacent faces of a cube.}
  \label{fig:threenonadj}
\end{figure}

{\bf Case 6:} A ring of four faces are replaced by tracks. We have the
drawing of Figure~\ref{fig:cubedrawings} (c). It is unique too. 

There are no other cases left. Thus a cube only has four confluent
drawings. 
\end{proof}

\begin{theorem}
The hypercube is non-confluent.
\end{theorem}

\begin{proof}
If we have a valid confluent drawing of the hypercube, and choose eight
of its vertices in the form of a cube, the portion of the drawing connecting these
vertices must be in one of the forms listed in the lemma above.
We consider the four possible drawings of this cube,
and attempt to add the other eight vertices (which also form a cube),
showing that each case leads to a contradiction.
Note that, among the edges of the first cube's drawing, only the ones
drawn as single edges can take part in confluent tracks with the remaining eight vertices.

{\bf Case 0:} In this drawing no faces are replaced.
Since the hypercube is non-planar, at least one of its faces must be replaced,
so we can always choose our first cube in such a way that this case does not occur.

{\bf Case 1:} Two adjacent faces of the cube are replaced, as in
Figure~\ref{fig:cubedrawings} (a).  If only two
adjacent faces $f_1$ and $f_2$ of the cube $C_1$ are replaced by
tracks, find a different cube $C_2$ sharing $f_1$ but not $f_2$ with
$C_1$. $C_2$ must have a second replaced face $f_3$ (it not possible
for a cube to have a confluent drawing with only one face replaced).
Either $f_1-f_3$ and $f_2-f_3$ are non-adjacent faces of the same
cube.  So if this case exists, we can find a different cube that is in
case 2 or case 3.

{\bf Case 3:} Two opposite faces of the cube are replaced, as in
Figure~\ref{fig:cubedrawings} (b).  In this
drawing, each face of the cube has only two non-track edges, each of
which can be crossed by at most one edge from the rest of the
graph. Because the other eight vertices of the graph form a cube which is $3$-connected,
any subset of these eight vertices has more than two edges connecting to
the complement of the subset. So putting any subset of these
vertices, other than the whole set, in a single face of the cube
drawing does not work.  Putting the whole set of the remaining vertices in a
single face of the cube drawing does not work either because there are
four vertices of the first cube outside that single face to be reached, and
only two of them can be reached across the two non-track edges.

{\bf Case 4:} A ring of four faces of the cube are replaced, as in
Figure~\ref{fig:cubedrawings} (c).   Edges between the other
eight vertices can not cross the tracks, so these vertices must all be placed
within a single face of the first cube's drawing.  However, these vertices would then be unable to connect to the four or more vertices of the first cube outside that face.

Since all cases fail, the $4$-dimensional hypercube is non-confluent.
\end{proof}

\section{Conclusions}
\label{sec:conclusions}

We introduce a new method of drawing non-planar graphs in planar way.  This 
can be very helpful for drawing graphs in the area of Software Visualization. 
Though we only show its applications on drawing function call graphs and 
object-interaction graphs, it is powerful for visualizing other kinds
of graphs too. 

\subsection*{Acknowledgments}
We would like to thank 
Jie Ren
and
Andr{\'e} van~der~Hoek
for supplying us with several examples of graphs used in software
visualization.
David Eppstein's research was supported by NSF grant CCR-9912338.

\bibliographystyle{abbrv}
\bibliography{conf,geom,geom-updates,goodrich} 
%
%
\balancecolumns
\end{document}